\documentclass{article}

\usepackage[preprint]{neurips_2025}

\usepackage[utf8]{inputenc} 
\usepackage[T1]{fontenc}    
\usepackage{hyperref}       
\usepackage{url}            
\usepackage{booktabs}       
\usepackage{amsfonts}       
\usepackage{nicefrac}       
\usepackage{microtype}      
\usepackage{xcolor}         
\usepackage{graphicx}
\usepackage{amsfonts}       
\usepackage{amsmath}        
\usepackage{float}

\title{NucEL: Single-Nucleotide ELECTRA-Style Genomic Pre-training for Efficient and Interpretable Representations}

\author{%
  Ke Ding \\
  John Curtin School of Medical Research\\
  Australian National University\\
  \texttt{Ke.Ding@anu.edu.au} \\
  \And
  Brian Parker \\
  School of Computing\\
  Australian National University\\
  \texttt{Brian.Parker@anu.edu.au} \\
  \And
  Jiayu Wen \\
  John Curtin School of Medical Research\\
  Australian National University\\
  \texttt{Jiayu.Wen@anu.edu.au} \\
}

\begin{document}

\maketitle

\begin{abstract}
  Pre-training large language models on genomic sequences has become a powerful approach for learning biologically meaningful representations. While masked language modeling (MLM)-based approaches, such as DNABERT and Nucleotide Transformer (NT), achieve strong performance, they are hindered by inefficiencies due to partial token supervision, pre-training/fine-tuning mismatches, and high computational costs. We introduce NucEL, the first ELECTRA-style pre-training framework for genomic foundation models, which overcomes these challenges. Through a discriminator network identifying tokens modified by a generator, NucEL achieves comprehensive token-level supervision across all sequence positions, thereby markedly improving training efficiency relative to the partial supervision of masked positions inherent in MLM frameworks. By integrating ModernBERT’s architectural advancements, including hybrid local-global attention and flash attention mechanisms, NucEL establishes an optimized BERT architecture for genomic sequence modeling. Unlike traditional methods that tokenize genomic sequences into 6-mers, NucEL implements single-nucleotide tokenization, enabling fine-grained resolution and improving both efficiency and interpretability. Pre-trained on the human genome only, NucEL achieves state-of-the-art performance on benchmark datasets across diverse downstream tasks in both human and non-human species, including regulatory element identification (e.g., promoters, enhancers), transcription factor binding prediction in human and mouse, open chromatin region classification, and histone modification profiles, surpassing MLM-based models of similar size and rivaling models 25 times larger, such as NT. Ablation studies provide critical insights into tokenization and masking strategies, optimizing ELECTRA-style pretraining for DNA sequences. Attention analyses reveal NucEL’s superior ability to capture biologically relevant sequence motifs compared to NT, offering valuable insights into its hierarchical learning process and regulatory element modeling capabilities. This work highlights the potential of ELECTRA-style pretraining as an efficient and effective strategy for advancing genomic representation learning with broad implications for future genomic research.
\end{abstract}

\section{Introduction}
Understanding genomic sequences is a fundamental challenge in biology, as DNA encodes the instructions for cellular function, development, and evolution.  Large language models (LLMs), which have transformed natural language processing (NLP), are now being adapted for biological sequences, showing remarkable success across genomics, transcriptomics, proteomics, metagenomics, and epigenomics \citep{RN34, RN70,  RN61, RN62}. Genomic language models (gLMs) are particularly promising for decoding the functional logic of genomes, revealing how DNA sequences are organized and interact to produce the complexity of biological systems.

Traditional deep learning methods for genomics, such as convolutional neural networks (CNNs) \citep{RN63} and recurrent neural networks (RNNs) \citep{RN64}, are limited: CNNs struggle with long-range dependencies in DNA sequences, while RNNs process sequences inefficiently. Transformer-based architectures overcome these shortcomings by leveraging pre-training on vast unlabeled genomic datasets to learn meaningful representations of functional elements such as promoters and enhancers, facilitating effective transfer learning for downstream tasks \citep{RN58}. However, existing transformer-based models, such as DNABERT \citep{RN35} and Nucleotide Transformer \citep{RN34}, rely on masked language modeling (MLM), which is computationally inefficient. MLM learns from only a small subset of the input—typically 15\% of masked tokens—underutilizing the majority of the sequence during training \citep{RN42}. Additionally, the pre-training (with masking) and fine-tuning (without masking) mismatch can impair transfer learning efficacy.

To address these limitations, we introduce NucEL, the first genomic foundation model to adopt the ELECTRA framework’s replaced token detection (RTD) objective \citep{RN42}. Unlike MLM, RTD employs a generator-discriminator architecture, enabling dense token-level supervision across all input positions and significantly boosting training efficiency. NucEL integrates ModernBERT’s architectural innovations \citep{RN40}, including hybrid local-global attention \citep{RN68} and flash attention mechanisms \citep{RN66}, to capture both short and long-range genomic dependencies effectively. Unlike k-mer \citep{RN34, RN35} or BPE tokenization \citep{RN36, RN39}, NucEL uses single-nucleotide tokenization, which may yield more tokens for long sequences but ensures fine-grained resolution and interpretability \citep{RN65}. Combining single-nucleotide resolution with RTD optimizes efficiency and precision: it preserves base-level detail for tasks like mutation analysis, while dense supervision and hybrid attention mitigate the computational cost of longer sequences and enhance long-range interaction modeling. We evaluate NucEL across a range of benchmark tasks—including promoter classification, transcription factor binding prediction, and epigenetic mark profiling—and demonstrate that it achieves state-of-the-art performance while using significantly fewer parameters than larger MLM-based models.
Our contributions include:
\begin{itemize}
    \item Introduced the first ELECTRA-style genomic foundation model, leveraging Replaced Token Detection (RTD) to enhance training efficiency over MLM-based approaches by utilizing all input tokens.
    \item Achieved state-of-the-art performance on genome foundation model benchmarks with significantly fewer parameters than MLM-based models.
    \item Utilized single-nucleotide resolution and fine-grained attention analyses to enable precise genomic modeling while enhancing interpretability by revealing complex biological representations.
    \item Integrated ModernBERT’s hybrid local-global attention and flash attention mechanisms to efficiently capture genomic dependencies while optimizing computational performance.

\end{itemize}

Our results demonstrate that NucEL provides a practical and efficient approach to learning biologically meaningful genomic representations, advancing genomic language modeling while reducing computational demands, thus making advanced genomic models more accessible to the broader research community.

\section{Related Work}
Genomic sequence modeling has evolved rapidly with deep learning, particularly through transformer-based models adapted from natural language processing. Here, we review key developments in genomic foundation models, pretraining objectives and tokenization strategies, positioning NucEL within this landscape.

\subsection{Genomic Language Models}

Transformer-based models have become a cornerstone of genomic representation learning. DNABERT~\citep{RN35} pioneered this approach by adapting BERT’s masked language modeling (MLM) to DNA sequences, using $k$-mer tokenization to predict masked tokens for tasks like promoter prediction. DNABERT-2~\citep{RN36} refined this with Byte Pair Encoding (BPE) tokenization and architectural improvements, enhancing performance on genomic benchmarks. The Nucleotide Transformer (NT) series~\citep{RN34} scaled this paradigm to billions of parameters, pretrained on multi-species datasets, demonstrating strong transfer learning capabilities. Additionally, ModernBERT~\citep{RN40} introduced architectural innovations such as hybrid local-global attention and flash attention mechanisms, improving transformer efficiency for sequence modeling. However, MLM-based models are limited by partial token supervision (only $\sim$15\% of tokens) and a pretraining-finetuning mismatch, reducing efficiency.

Alternatives to transformers include state-space models (SSMs) like HyenaDNA~\citep{RN37} and Caduceus~\citep{RN38}, which use single-nucleotide resolution and reverse-complement awareness for efficient long-sequence modeling. Convolutional neural networks (CNNs), such as DeepBind~\citep{RN54} and Enformer~\citep{RN56}, remain relevant for capturing local motifs but struggle with long-range dependencies compared to transformers.

\subsection{Pretraining Objectives}

Pretraining objectives shape the efficiency and effectiveness of genomic models. MLM, used in DNABERT~\citep{RN35} and NT~\citep{RN34}, masks a subset of tokens for prediction, but its sparse supervision leaves most of the forward pass unutilized. Next-Sentence Prediction (NSP), used alongside with MLM in BERT~\citep{RN58}, is rarely adopted in genomics due to DNA’s lack of paragraph-like structure, offering limited benefits~\citep{RN52}. In contrast, Replaced Token Detection (RTD), introduced by ELECTRA~\citep{RN42}, uses a generator-discriminator framework to provide dense supervision across all tokens, significantly improving efficiency.  RTD has proven effective in NLP (e.g., DeBERTa-v3~\citep{RN41}), but its potential in genomics remains untapped, motivating our NucEL’s approach.

\subsection{Tokenization Strategies}

Tokenization critically shapes a model’s ability to represent genomic sequences. $K$-mer tokenization, as used in DNABERT~\citep{RN35}, compresses sequences by grouping nucleotides into $k$-length substrings but fractures single-nucleotide context, exponentially inflating vocabulary size and diluting base-level detail. Byte Pair Encoding (BPE), adopted by DNABERT-2~\citep{RN36} and GROVER~\citep{RN39}, learns variable-length subwords to balance efficiency and motif capture but can obscure single-nucleotide variations and introduce inconsistencies for similar sequences. Single-nucleotide tokenization, as in HyenaDNA~\citep{RN37} and Caduceus~\citep{RN38}, preserves base-level resolution and avoids vocabulary explosion, making it ideal for tasks requiring fine-grained precision, such as modeling point mutations and identifying regulatory motifs such as transcription factor binding motifs.

\subsection{Positioning NucEL}

While MLM-based models perform well on downstream tasks, their inefficiencies in supervision and computational demands drive the need for alternatives. NucEL introduces the first ELECTRA-style framework for genomic modeling, leveraging RTD to achieve dense supervision and superior sample efficiency compared to MLM-based approaches. By integrating ModernBERT’s hybrid attention and flash attention mechanisms~\citep{RN40} and adopting single-nucleotide tokenization, NucEL maximizes interpretability and efficiency. This enables state-of-the-art performance with fewer parameters, positioning NucEL as a scalable and effective solution for genomic representation learning.

\section{Methods}
This section describes NucEL’s training strategies, model architecture, pre-training data and tokenization, optimization, and fine-tuning approach.

\subsection{Training Strategies}
NucEL uses a RTD objective, training two transformer encoders in a generator-discriminator framework. The generator, a smaller transformer, performs MLM to predict masked tokens, while the discriminator, a larger transformer, identifies whether tokens are original or replaced by the generator. This provides dense supervision across all tokens, enhancing sample efficiency. Quantitatively, RTD supervises all \(N\) tokens in a sequence of length \(N\), unlike MLM’s sparse supervision of \(r \cdot N\) tokens (e.g., \(r = 0.15\)), yielding a supervision ratio of approximately 6.67. The total loss combines the generator’s MLM loss and the discriminator’s RTD loss, weighted by \(\lambda = 50.0\):
\begin{equation}
\mathcal{L}_{\text{total}} = \mathcal{L}_{\text{generator}} + \lambda \cdot \mathcal{L}_{\text{discriminator}},
\end{equation}
where \(\mathcal{L}_{\text{generator}}\) is the cross-entropy over masked tokens, and \(\mathcal{L}_{\text{discriminator}}\) is the binary cross-entropy over all tokens.

For a sequence \(\mathcal{X} = \{x_1, x_2, \ldots, x_N\}\) with masked positions \(\mathcal{M} \subseteq \{1, 2, \ldots, N\}\), the generator loss is:
\begin{equation}
\mathcal{L}_{\text{generator}} = -\frac{1}{|\mathcal{M}|} \sum_{i \in \mathcal{M}} \log P_G(x_i | \mathcal{X}_{\text{masked}}),
\end{equation}
and the discriminator loss is:
\begin{equation}
\mathcal{L}_{\text{discriminator}} = -\frac{1}{N} \sum_{i=1}^N \left[ y_i \log P_D(y_i = 1 | \mathcal{X}') + (1 - y_i) \log (1 - P_D(y_i = 1 | \mathcal{X}')) \right],
\end{equation}
where \(\mathcal{X}'\) is the sequence with masked tokens replaced by the generator’s predictions, and \(y_i = 1\) if \(x'_i = x_i\), else 0.

\subsection{Model Architecture}

\begin{figure}
\centering
  \includegraphics[width=0.8\textwidth]{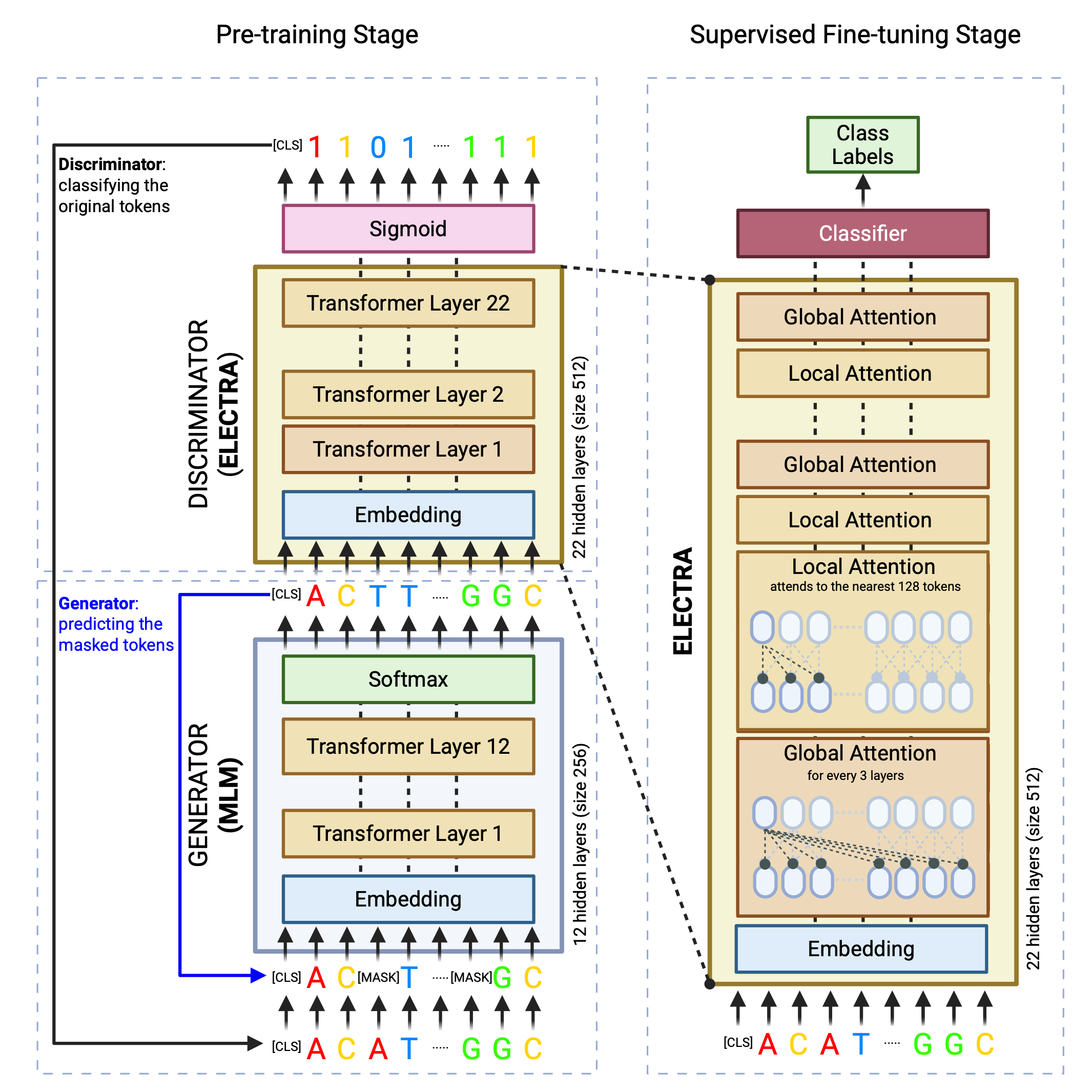}
  \caption{\textbf{Overview of NucEL the Electra-style Genomic Pre-training Framework.} Illustrates the two-stage ELECTRA-style pre-training framework for NucEL, depicting the generator-discriminator architecture during the pre-training stage and the supervised fine-tuning stage for downstream genomic tasks.}
  \label{fig:model_arch}
  
\end{figure}

NucEL’s generator-discriminator framework is shown in Figure~\ref{fig:model_arch}. The generator comprises 11 transformer layers, a hidden size of 256, and 8 attention heads. The discriminator, NucEL’s primary model, has 22 layers, a hidden size of 512, and 16 attention heads. Both use a hybrid attention mechanism, combining local attention windows (128 tokens) with global attention every third layer, balancing efficiency and long-range dependency modeling (see details in Appendix).

\subsection{Pre-training Data and Tokenization}
NucEL was pre-trained on the human genome (GRCh38/Hg38) using single-nucleotide tokenization, where each base (A, C, G, T) is a distinct token. Sequences were extracted using an overlapping sliding window (1224 bp, 100 bp overlap), with 1024 bp segments randomly sampled during training. The vocabulary consists of 27 tokens: 4 nucleotides, 7 special tokens ([PAD], [UNK], [SEP], [CLS], [MASK], [BOS], [EOS]), and 16 reserved for future extensions (see Appendix). Ambiguous bases (N) are mapped to [UNK] or omitted, though rare.

Single-nucleotide tokenization preserves base-level precision but increases sequence length compared to alternatives like k-mer (k=6) and Byte Pair Encoding (BPE). For a sequence of length \(L\), the tokenized length for single-nucleotide tokenization is \(N_{\text{1-mer}} = L\), while for k-mer tokenization with overlapping windows, it is \(N_{\text{k-mer}} = L - k + 1\) (e.g., for \(k=6\)), and for BPE, it is approximately \(N_{\text{BPE}} \approx \alpha L\), where \(0.1 < \alpha < 0.5\) depending on the learned vocabulary. The vocabulary size for single-nucleotide tokenization is \(|\mathcal{V}_{\text{1-mer}}| = 27\), significantly smaller than \(|\mathcal{V}_{\text{k-mer}}| = 4^k\) (e.g., \(4^6 = 4096\) for \(k=6\)) for k-mer tokenization, and \(|\mathcal{V}_{\text{BPE}}| \approx 5000 \text{ to } 8000\) for BPE. While single-nucleotide tokenization increases sequence length, this is mitigated by NucEL’s local attention mechanism, which reduces computational complexity while preserving base-level precision essential for fine-grained genomic tasks.

\subsection{Pre-training and Fine-tuning Details}
NucEL was trained using the AdamW optimizer (learning rate \(1 \times 10^{-4}\), \(\beta_1 = 0.9\), \(\beta_2 = 0.999\)) for 50 epochs, with a global batch size of 192 and mixed precision (FP16) on 8 NVIDIA A100 GPUs. Training stability was ensured by a 1000-step warmup schedule and a maximum gradient norm of 1.0 (see Appendix).
For downstream tasks, a linear layer was added atop the discriminator’s [CLS] token output, and the entire model was fine-tuned end-to-end.

\section{Experiments}
We evaluated NucEL on three human genome-focused benchmark suites: Genome Understanding Evaluation (GUE) \citet{RN36}, Genomic Benchmarks (GB) \citet{RN69}, and Nucleotide Transformer (NT) \citet{RN34} tasks. These cover diverse genomic tasks, including gene regulation, epigenomics, and sequence classification (see Appendix for details). Performance was assessed using the Matthews Correlation Coefficient (MCC) for all benchmarks ensuring robust evaluation across imbalanced datasets. In addition, accuracy was reported for GB tasks.
 
\subsection{Evaluating Regulatory Sequence Recognition on GUE Benchmark}
We first evaluate NucEL on the human subset of the GUE benchmark, including transcription factor binding (TF-H), promoter detection (PD), core promoter detection (CPD), and splice site prediction (SSP). Following DNABERT2's protocol~\citep{RN36}, we report average MCC across three random seeds, selecting the model with the lowest validation loss. Compared models include DNABERT (DB1, 89M parameters), DNABERT-2 (DB2, 117M), and Nucleotide Transformer variants (500M and 2.5B parameters, trained on human or multi-species data)\citep{RN36}. As shown in Table\ref{tab:gue_result}, NucEL (93M parameters) demonstrates competitive performance across human genomic tasks, achieving top-1 results on CPD and SSP tasks.

We further evaluate NucEL's generalization ability across different species to assess its cross-domain transferability. Despite being trained exclusively on human genomic data, NucEL demonstrates remarkable performance on diverse genomic tasks across mouse (Transcription Factor binding, TF-M), yeast (Epigenetic Marks Prediction, EMP), and viral genomes (Covid Variant Classification, CVC), as shown in Table~\ref{tab:gue_result}. Notably, NucEL outperforms models trained on multi-species data while NT-multi contains 25 times more parameters than NucEL. The model achieves the highest overall average performance of 75.16 across all seven genomic tasks and sets top-2 results on 6 of 7 cross-species tasks, demonstrating that fundamental genomic patterns learned from human data can effectively generalize to other species. This cross-species transferability, combined with its parameter efficiency, highlights NucEL's ability to capture universal genomic representations that transcend species boundaries.

\begin{table}[htbp]
\centering
\fontsize{9pt}{9pt}\selectfont
\caption{Performance comparison on GUE (MCC for all tasks except CVC which uses F1; Averaged over 3 Seeds; \textbf{Best}, \underline{Second-Best})}
\setlength{\tabcolsep}{10pt}
\begin{tabular}{lcccc}
\hline
& \multicolumn{4}{c}{\textbf{Human Tasks}} \\ 
\cline{2-5}
\noalign{\vskip 0.8ex}
\textbf{Model} & \textbf{TF-H} & \textbf{PD} & \textbf{CPD} & \textbf{SSP} \\
\hline
\noalign{\vskip 0.8ex}
DNABERT (3-mer) & 64.43 & 84.63 & \underline{72.96} & 84.14 \\
DNABERT (4-mer) & 64.41 & 82.99 & 71.10 & 84.05 \\
DNABERT (5-mer) & 50.46 & 84.04 & 72.03 & 84.02 \\
DNABERT (6-mer) & 64.17 & 81.70 & 71.81 & 84.07 \\
NT-500M-human & 50.82 & 85.51 & 66.54 & 79.71 \\
NT-500M-1000g & 58.92 & 86.58 & 69.13 & 80.97 \\
NT-2500M-1000g & 61.99 & 86.61 & 68.17 & 85.78 \\
NT-2500M-multi & 63.32 & \textbf{88.14} & 71.62 & \underline{89.36} \\
DNABERT-2 (117M) & \textbf{70.10} & 84.21 & 70.52 & 84.99 \\
\hline
\noalign{\vskip 0.8ex}
\textbf{NucEL (93M)} & \underline{67.64} & \underline{87.10} & \textbf{75.13} & \textbf{90.30} \\
\hline
& \multicolumn{3}{c}{\textbf{Other Species}} &\\ 
\cline{2-4}
\noalign{\vskip 0.8ex}
\textbf{Model} & \textbf{TF-M} & \textbf{EMP} & \textbf{CVC} & \textbf{Average} \\
& \textbf{(Mouse)} & \textbf{(Yeast)} & \textbf{(Virus)} & \textbf{Performance} \\
\hline
\noalign{\vskip 0.8ex}
DNABERT (3-mer) & 57.73 & 49.54 & 62.23 & 67.95 \\
DNABERT (4-mer) & 59.58 & 48.59 & 59.87 & 67.23 \\
DNABERT (5-mer) & 54.85 & 48.62 & 63.64 & 65.38 \\
DNABERT (6-mer) & 56.43 & 49.10 & 55.50 & 66.11 \\
NT-500M-human & 45.24 & 45.35 & 57.13 & 61.47 \\
NT-500M-1000g & 49.31 & 47.68 & 52.06 & 63.52 \\
NT-2500M-1000g & 56.82 & 50.86 & 66.73 & 68.14 \\
NT-2500M-multi & 67.01 & 58.06 & \textbf{73.04} & \underline{72.94} \\
DNABERT-2 (117M) & \underline{67.99} & 55.98 & \underline{71.02} & 72.11 \\
\hline
\noalign{\vskip 0.8ex}
\textbf{NucEL (93M)} & \textbf{70.62} & \textbf{65.01} & 70.29 & \textbf{75.16} \\
\hline
\end{tabular}
\label{tab:gue_result}
\end{table}

\subsection{Assessing Regulatory DNA Classification on GB Benchmark}
We assessed NucEL on seven human-centric regulatory sequence classification tasks from the Genomic Benchmarks dataset~\citep{RN69}, evaluating its ability to distinguish functional DNA classes across diverse sequence lengths. Adopting Caduceus’s protocol~\citep{RN38}, we report average accuracy and standard deviation across five random seeds, selecting the model with the lowest validation loss. Compared models include HyenaDNA, Caduceus variants (Ph and PS), and Nucleotide Transformer 2 (NT2-100M)~\citep{RN38}. Table~\ref{tab:genomic_benchmark_results} shows NucEL achieves the highest average accuracy of 89.9\%, outperforming NT2-100M (89.0\%) and Caduceus-Ph (88.2\%), with state-of-the-art results on four tasks and second-best on two, underscoring its robustness in regulatory DNA classification.

\begin{table}[h]
\centering
\fontsize{9pt}{9pt}\selectfont
\caption{Performance comparison on GB (Accuracy; Averaged over 5 Seeds with Standard Deviation; \textbf{Best},  \underline{Second-Best})}
\label{tab:genomic_benchmark_results}
\begin{tabular}{@{}lccccc@{}}
\toprule
Genomic Benchmark Tasks & HyenaDNA & Caduceus-Ph & Caduceus-PS & NT2-100m & \textbf{NucEL} \\
\midrule
Coding vs. Intergenic & 0.904 {\tiny $\pm$0.005} & 0.915 {\tiny $\pm$0.003} & 0.910 {\tiny $\pm$0.003} & \textbf{0.950} {\tiny $\pm$0.002} & \underline{0.941} {\tiny $\pm$0.001} \\
Human vs. Worm & 0.964 {\tiny $\pm$0.002} & \underline{0.973} {\tiny $\pm$0.001} & 0.968 {\tiny $\pm$0.002} & 0.972 {\tiny $\pm$0.001} & \textbf{0.975} {\tiny $\pm$0.001} \\
Human Enhancers Cohn & 0.729 {\tiny $\pm$0.014} & \textbf{0.747} {\tiny $\pm$0.004} & \underline{0.745} {\tiny $\pm$0.007} & 0.736 {\tiny $\pm$0.004} & 0.735 {\tiny $\pm$0.008} \\
Human Enhancer Ensembl & 0.849 {\tiny $\pm$0.006} & 0.893 {\tiny $\pm$0.008} & 0.900 {\tiny $\pm$0.006} & \underline{0.935} {\tiny $\pm$0.001} & \textbf{0.940} {\tiny $\pm$0.001} \\
Human Regulatory & 0.869 {\tiny $\pm$0.012} & 0.872 {\tiny $\pm$0.011} & 0.873 {\tiny $\pm$0.007} & \textbf{0.941} {\tiny $\pm$0.001} & \underline{0.932} {\tiny $\pm$0.002} \\
Human OCR Ensembl & 0.783 {\tiny $\pm$0.007} & \textbf{0.828} {\tiny $\pm$0.006} & \underline{0.818} {\tiny $\pm$0.006} & 0.776 {\tiny $\pm$0.002} & 0.794 {\tiny $\pm$0.002} \\
Human NonTATA Promoters & 0.944 {\tiny $\pm$0.002} & \underline{0.946} {\tiny $\pm$0.007} & 0.945 {\tiny $\pm$0.010} & 0.920 {\tiny $\pm$0.003} & \textbf{0.973} {\tiny $\pm$0.001} \\
\midrule
Average & 0.863 & 0.882 & 0.880 & \underline{0.890} & \textbf{0.899} \\
\bottomrule
\end{tabular}
\end{table}

\subsection{Profiling Genomic Element Prediction on NT Benchmark}
We evaluated NucEL on the revised Nucleotide Transformer benchmark~\citep{RN34}, encompassing histone marker prediction, regulatory element identification, and splice site detection. Following~\citet{RN34}, we report average MCC across ten random seeds. Table~\ref{tab:nt2_results} shows NucEL (93M parameters) outperforms similarly sized models like DNABERT2 (117M) and NT2-Multi (100M), achieving state-of-the-art on 11 of 18 tasks. Its average MCC of 0.664 matches NT2-Multi (500M parameters) and slightly exceeds NT-Multi (2.5B parameters, 0.661), despite using only human genome data for pre-training, unlike the multi-species data used by DNABERT2 and NT2. NucEL’s competitive performance on chromatin profiles (e.g., histone markers) is detailed in the Appendix.  With five times fewer parameters than NT2-500M and 27 times fewer than NT-Multi 2.5B, NucEL demonstrates exceptional efficiency and performance.

\begin{table}[h]
\centering
\fontsize{10pt}{10pt}\selectfont
\caption{Performance comparison on NT (MCC; Averaged over 10 Seeds with Standard Deviation; \textbf{Best},  \underline{Second-Best})}
\label{tab:nt2_results}
\begin{tabular}{@{}lcccc@{}}
\toprule
Task & DNABERT2 & NT-HumanRef & NT2-Multi & NucEL\\
 & (117M) & (500M) & (100M) & (93M)\\
\midrule
H2AFZ & 0.49 {\tiny $\pm$0.013} & 0.465 {\tiny $\pm$0.011} & \underline{0.492} {\tiny $\pm$0.012} & \textbf{0.508} {\tiny $\pm$0.007} \\
H3K27ac & \textbf{0.491} {\tiny $\pm$0.01} & 0.457 {\tiny $\pm$0.01} & \underline{0.487} {\tiny $\pm$0.016} & 0.470 {\tiny $\pm$0.006} \\
H3K27me3 & \underline{0.599} {\tiny $\pm$0.01} & 0.589 {\tiny $\pm$0.009} & 0.595 {\tiny $\pm$0.013} & \textbf{0.611} {\tiny $\pm$0.016} \\
H3K36me3 & \textbf{0.637} {\tiny $\pm$0.007} & 0.594 {\tiny $\pm$0.004} & \underline{0.617} {\tiny $\pm$0.006} & 0.584 {\tiny $\pm$0.007} \\
H3K4me1 & \textbf{0.49} {\tiny $\pm$0.008} & 0.468 {\tiny $\pm$0.007} & \underline{0.485} {\tiny $\pm$0.011} & 0.480 {\tiny $\pm$0.011} \\
H3K4me2 & \textbf{0.558} {\tiny $\pm$0.013} & 0.527 {\tiny $\pm$0.011} & \underline{0.551} {\tiny $\pm$0.01} & 0.546 {\tiny $\pm$0.010} \\
H3K4me3 & \textbf{0.646} {\tiny $\pm$0.008} & 0.622 {\tiny $\pm$0.013} & \underline{0.633} {\tiny $\pm$0.015} & 0.621 {\tiny $\pm$0.019} \\
H3K9ac & \textbf{0.564} {\tiny $\pm$0.013} & 0.524 {\tiny $\pm$0.013} & 0.538 {\tiny $\pm$0.015} & \underline{0.550} {\tiny $\pm$0.011} \\
H3K9me3 & 0.443 {\tiny $\pm$0.025} & 0.433 {\tiny $\pm$0.009} & \underline{0.445} {\tiny $\pm$0.017} & \textbf{0.474} {\tiny $\pm$0.013} \\
H4K20me1 & \textbf{0.655} {\tiny $\pm$0.011} & 0.634 {\tiny $\pm$0.013} & \underline{0.648} {\tiny $\pm$0.008} & 0.640 {\tiny $\pm$0.009} \\
Enhancer & \underline{0.517} {\tiny $\pm$0.011} & 0.515 {\tiny $\pm$0.019} & 0.507 {\tiny $\pm$0.009} & \textbf{0.578} {\tiny $\pm$0.005} \\
Enhancer types & 0.476 {\tiny $\pm$0.009} & \underline{0.477} {\tiny $\pm$0.014} & 0.465 {\tiny $\pm$0.009} & \textbf{0.536} {\tiny $\pm$0.002} \\
Promoter all & \underline{0.754} {\tiny $\pm$0.009} & 0.734 {\tiny $\pm$0.013} & 0.753 {\tiny $\pm$0.005} & \textbf{0.762} {\tiny $\pm$0.005} \\
Promoter non-TATA & 0.762 {\tiny $\pm$0.006} & 0.738 {\tiny $\pm$0.008} & \underline{0.766} {\tiny $\pm$0.014} & \textbf{0.769} {\tiny $\pm$0.007} \\
Promoter TATA & 0.784 {\tiny $\pm$0.036} & \underline{0.831} {\tiny $\pm$0.022} & 0.826 {\tiny $\pm$0.019} & \textbf{0.922} {\tiny $\pm$0.004} \\
Splice acceptor & 0.837 {\tiny $\pm$0.006} & 0.941 {\tiny $\pm$0.004} & \underline{0.947} {\tiny $\pm$0.003} & \textbf{0.960} {\tiny $\pm$0.003} \\
Splice site all & 0.835 {\tiny $\pm$0.005} & 0.939 {\tiny $\pm$0.003} & \underline{0.960} {\tiny $\pm$0.005} & \textbf{0.962} {\tiny $\pm$0.004} \\
Splice donor & 0.861 {\tiny $\pm$0.004} & \underline{0.952} {\tiny $\pm$0.003} & 0.947 {\tiny $\pm$0.008} & \textbf{0.970} {\tiny $\pm$0.002} \\
\midrule
Average & 0.628 & 0.634 & \underline{0.651} & \textbf{0.664} \\
\bottomrule
\end{tabular}
\end{table}

\subsection{Impact of Tokenization Strategies}
\begin{figure}[h]
  \centering
  \includegraphics[width=1.0\textwidth]{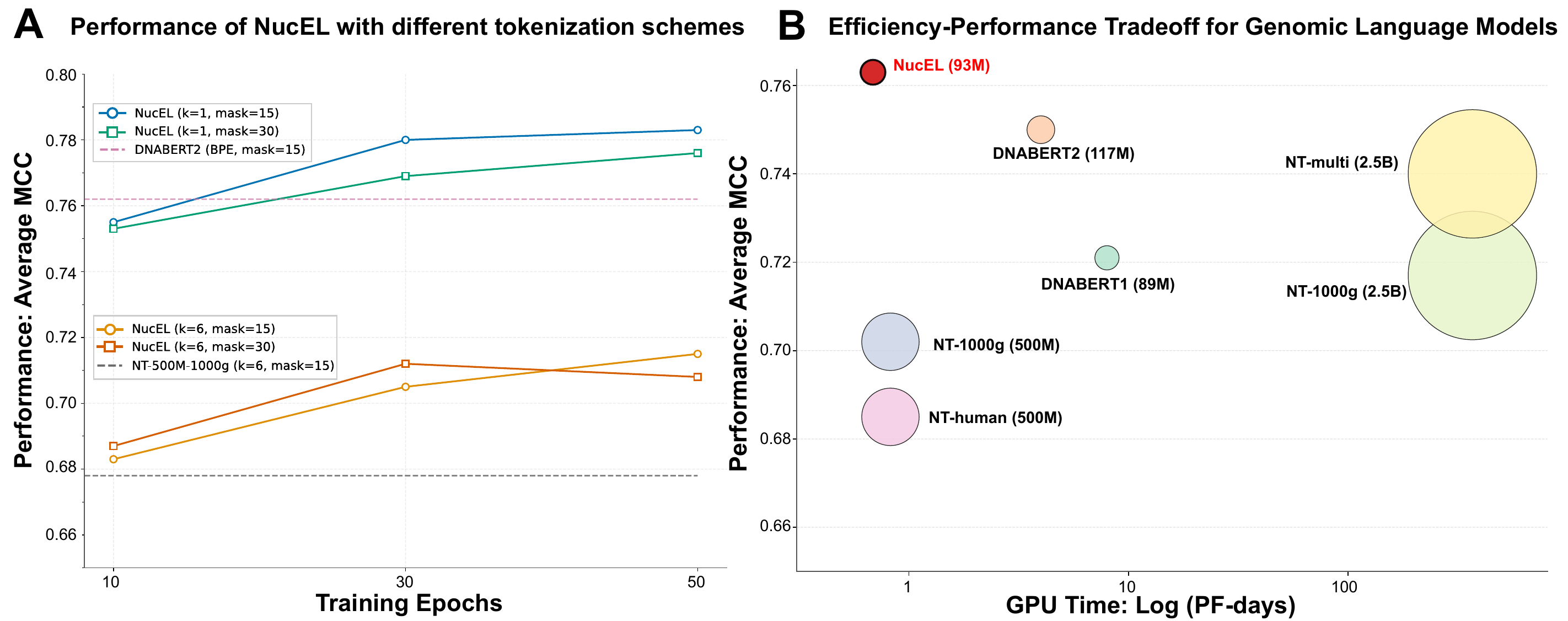}
  \caption{\textbf{(A) Performance Across Tokenization Schemes.} Performance of NucEL with different tokenization schemes (single-nucleotide, k-mer with k=6, and BPE) across training epochs on human GUE datasets, highlighting the superior representation learning of single-nucleotide tokenization. \textbf{(B) Efficiency-Performance Tradeoff for Genomic Language Models.} Efficiency-performance tradeoff on human GUE tasks, with GPU time (petaflop-days, log scale) on the x-axis, average MCC on the y-axis, and bubble size representing model parameter count. NucEL-93M (in red) achieves a strong balance, outperforming larger models like NT-multi-2.5B.}
  \label{fig:efficiency_figure}
\end{figure}

Tokenization significantly impacts genomic language model performance and efficiency. Figure~\ref{fig:efficiency_figure}(A) compares NucEL’s performance across tokenization schemes—single-nucleotide (\(k=1\)), k-mer (\(k=6\)), and Byte Pair Encoding (BPE)—over training epochs on GUE datasets. Single-nucleotide tokenization outperforms k-mer and BPE, preserving base-level detail critical for genomic tasks. K-mer tokenization, as used in NT-500M-1000g, reduces sequence length but dilutes single-nucleotide context, compromising performance. BPE, employed by DNABERT2, improves computational efficiency but obscures single-nucleotide variations and risks inconsistent tokenization for similar sequences due to mutation-induced boundary shifts. Paired with ELECTRA’s RTD objective, single-nucleotide tokenization enables efficient learning across all sequence positions, yielding superior representations.

\subsection{Efficiency-Performance Tradeoff Analysis}
We analyzed NucEL’s efficiency-performance tradeoff against BERT-based genomic models, focusing on training time, performance (MCC), and model size. Figure~\ref{fig:efficiency_figure}(B) illustrates this tradeoff on GUE tasks, with GPU time (petaflop-days, log scale) on the x-axis, average MCC on the y-axis, and bubble size representing parameter count. NucEL-93M (red) achieves a superior balance, outperforming larger models like NT-multi-2.5B and NT-1000g-2.5B, which are approximately 25 times larger in parameter count and require over 100 times more computational resources. Compared to models with similar computational demands (e.g., NT-1000g-500M, NT-human-500M), NucEL delivers over 10\% higher MCC. This highlights the ELECTRA-style framework’s ability to provide robust genomic understanding with significantly reduced computational and parameter requirements.

\subsection{Embedding Analysis by Gene Biotype}
\begin{figure}[h]
  \centering
    \includegraphics[width=1\textwidth]{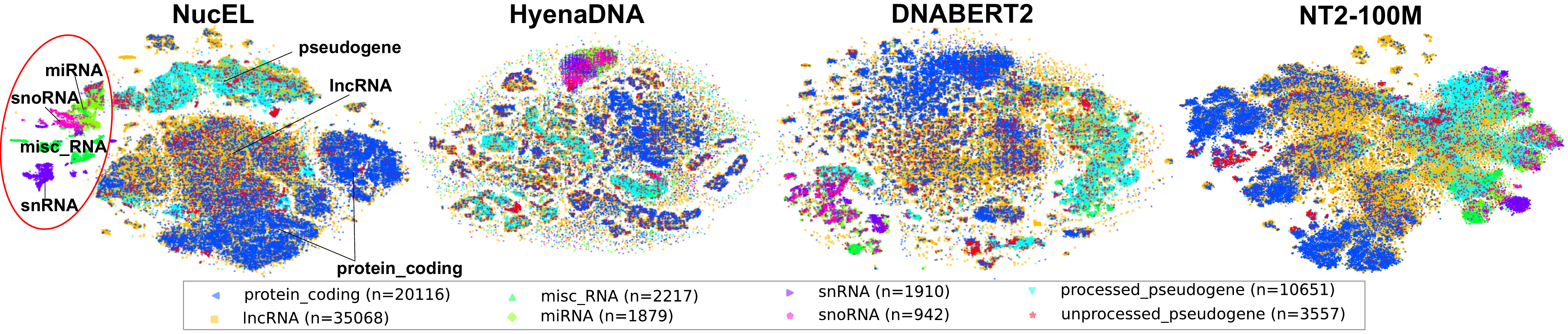}
  \caption{\textbf{t-SNE Visualization of Gene Biotype Embeddings.} t-SNE visualizations of embeddings generated by NucEL, HyenaDNA, DNABERT2, and NT2-100M for the prevalent biotypes, demonstrating NucEL’s superior separation.}
  \label{fig:tsme_embedding}
\end{figure}

To assess NucEL’s ability to capture biological signals, we extracted sequences from the most prevalent Ensembl biotypes and generated embeddings using NucEL, HyenaDNA (small 32K), DNABERT2 (117M), and NT2-100M. We visualized embeddings using t-SNE (Figure~\ref{fig:tsme_embedding}) and trained an XGBoost classifier to predict biotypes, evaluating performance with macro, micro, and weighted F1 scores (Table~\ref{tab:embedding_quality}). NucEL achieved the highest F1 scores across all metrics, demonstrating superior separation of biotype classes, particularly for small RNAs (red circle in Figure~\ref{fig:tsme_embedding}). Notably, lincRNA and protein-coding genes, which are often hard to distinguish due to sequence similarities and shared exon-intron structures, were effectively clustered together in NucEL’s embeddings, highlighting its robust representation of complex biological relationships compared to competing models.

\begin{table}[h]
\centering
\caption{Embedding quality Weighted F1 classification score on biotypes.}
\label{tab:embedding_quality}
\begin{tabular}{lccc}
\textsc{Model} & \textsc{Macro F1} & \textsc{Micro F1} & \textsc{Weighted F1} \\
\hline
HyenaDNA & 0.6882 & 0.7519 & 0.7377 \\
NT2-100m & 0.6325 & 0.7579 & 0.7470 \\
DNABERT2 & 0.6398 & 0.7487 & 0.7358 \\
\hline
\textbf{NucEL} & \textbf{0.7306} & \textbf{0.7603} & \textbf{0.7490} \\
\hline
\end{tabular}
\end{table}

\subsection{Visualizing Attention and Model Interpretability}

Predictive models for genomic sequences aim to uncover novel biological insights through interpretable analyses, surpassing traditional probabilistic enrichment studies by capturing complex interactions between regulatory elements. To evaluate model interpretability, we designed a synthetic “motif-order” classification task featuring two randomly placed motifs, A and B, where sequences with Motif A before B are positive and B before A are negative. This task mimics biological scenarios, such as RNA-binding proteins binding in a specific order to activate cis-regulatory elements. The motifs share identical base composition and are separated by a random gap, forcing the model to learn long-range interactions rather than relying on simple motif detection. See Appendix for detailed methods and results.

\begin{figure}[h]
\centering
\includegraphics[width=0.8\textwidth]{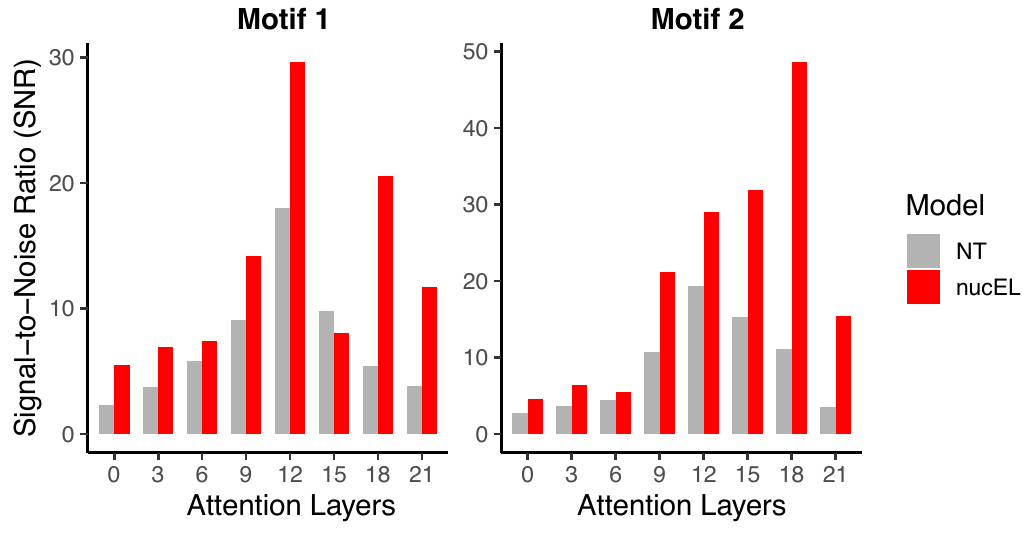}
\caption{Signal-to-Noise Ratio (SNR) Comparison: Comparison of SNR for attention weights in NucEL and NT2-100M across global attention layers on the motif-order classification task, showing NucEL’s superior performance in detecting Motif A and Motif B with higher SNR values, indicating clearer motif identification and reduced background noise compared to NT2-100M.}
\label{fig:snr_comparison}
\end{figure}

We compared NucEL and NT2-100M on this motif-order task, with both models fine-tuned to near-perfect accuracy. Despite similar predictive performance, their interpretability differed substantially. We extracted attention weights from the global attention layers (layers 0, 3, 6, 9, 12, 15, 18, and 21) and computed averaged attention maps across 100 test sequences aligned at the motif positions. Figure~\ref{fig:snr_comparison} shows the signal-to-noise ratio (SNR) at true motif locations, quantifying how strongly attention signals rise above background noise. NucEL consistently exhibits higher SNRs across global layers, indicating more effective localization of informative features. Attention heatmaps (see Appendix) further reveal that NucEL focuses sharply on the embedded motif regions, while NT2-100M displays more diffuse attention with elevated background activity. Quantitatively, NucEL achieves substantial improvements in maximum SNR—65\% higher for Motif A and 152\% higher for Motif B—highlighting its superior capacity to distinguish motif signals from noise. Since high SNR is critical for minimizing false discovery rates in regulatory motif identification, these findings underscore NucEL’s strength in fine-grained genomic modeling and its ability to capture long-range regulatory dependencies with greater precision and interpretability than existing models.

\section{Limitations}

While NucEL achieves state-of-the-art performance on human-genome benchmarks and demonstrates strong zero-shot generalization to non-human species (e.g., mouse, yeast, virus), its current pretraining remains confined to the human genome. Expanding pretraining to include diverse taxonomic groups (e.g., invertebrates, microbes) could enhance cross-species transferability, particularly for clade-specific regulatory elements or non-conserved motifs. This is a promising direction for future work. 

\section{Conclusion}
This study introduces NucEL, the first ELECTRA-style pre-training framework for genomic sequences, shifting from the conventional masked language modeling (MLM) paradigm. Extensive experiments demonstrate that NucEL’s replaced token detection (RTD) approach combined with single-nucleotide tokenization surpasses MLM, offering superior efficiency, accuracy, and interpretability in genomic representation learning. NucEL’s innovations include: (1) an ELECTRA-style generator-discriminator architecture for dense token-level supervision; (2) single-nucleotide tokenization, enabling precise modeling of fine-grained genomic features; and (3) ModernBERT’s hybrid attention mechanisms, capturing both local and global genomic dependencies. Combining single-nucleotide resolution with RTD enhances efficiency and precision, preserving base-level detail while dense supervision and hybrid attention mitigate computational costs. These advancements allow NucEL to achieve state-of-the-art performance across diverse genomic benchmarks with fewer parameters than competing models. Enhanced by fine-grained tokenization and attention mechanisms, NucEL provides clear insights into genomic features driving predictions, improving model interpretability. This work advances genomic language modeling by delivering a computationally efficient and interpretable tool, demonstrating the ability of new pre-training strategies to extract additional information in genomic analysis.

\begin{ack}
The authors acknowledge computational resources provided by the National Computational Infrastructure (NCI), the Argonne National Laboratory, and Google Cloud. This work was supported in part by the Talo Scholarship and the Talo Innovative Grant, provided by Taiyang Zhang and Loong Wang. The authors also acknowledge NCI for the HPC-AI Talent Program. We are especially thankful to Dr. Jingbo Wang and Dr. Arvind Ramanathan for their valuable discussions and constructive suggestions, which have contributed substantially to the development of this work.

\end{ack}

\newpage
\immediate\write18{cp \jobname.bbl visible.bbl}
\bibliographystyle{plainnat}  
\bibliography{reference}      

\end{document}